\documentclass{moriond}

\def\be{\begin{equation}}
\def\ee{\end{equation}}
\def\bea{\begin{eqnarray}}
\def\eea{\end{eqnarray}}

\usepackage[force]{feynmp-auto}

\usepackage{bclogo}
\usepackage{pifont}
\usepackage{amsmath}

\newcommand{\Photo}{}

\begin{document}
\vspace*{4cm}
\title{$S$-matrix approach to general gravity and
  beyond\footnote{IPhT-T21/019, CERN-TH-2021-062\\
    Contribution to the 2021 Gravitation session of the 55th Rencontres de
    Moriond  March 9-11, 2021.
}}

\author{ Pierre Vanhove }

\address{Institut de Physique Theorique, Universit\'e Paris-Saclay,
CEA, CNRS, F-91191 Gif-sur-Yvette Cedex, France\\
  National Research University Higher School of
  Economics, Russian Federation\\
  Theoretical Physics Department, CERN, 1211 Geneva 23, Switzerland}
\maketitle\abstracts{
In this text we outline the motivation for developping a quantum
$S$-matrix approach for the classical gravitational two-body
scattering. As an application we briefly present the derivation of black-hole metrics in
various dimensions.}

\section{Motiviation} 

The detection of the first gravitational wave signal has opened 
an area of \emph{precision} gravity.
The measurement of the gravitational wave signal is a formidable
window on gravity at various scales, involving the dynamics of black
holes. This gives precious information on Einstein gravity  and
possibly gravitational physics beyond it.
Ultimately, this would tell us how good we understand gravity both in
the weak and strong coupling regimes.
 The binary mergers detected until now by the LIGO-Virgo~\cite{Abbott:2016blz,TheLIGOScientific:2017qsa}
 collaboration are clean
  sources of gravitational waves and the gravitational-wave signal is currently
modelled by general relativity in vacuum, at accuracy close to second
order in the mass ratio parameter.
The next generation gravitational-wave detectors, such as the Laser Interferometer Space Antenna (LISA) will be
sensitive to the environement of the sources and the signal will be more ``dirty''~\cite{Barausse:2020rsu}.
The lack of clear
predictions for nonlinearities (from the accretion disk for instance) in the post-merger phase means that these could be confused with modifications of the
signal predicted by theories beyond general relativity.

For this we need to produce accurate theoretical gravitational
waveform templates. We have to clarify how much can be understood from exact
    theoretical computations. We have to answer 
    the questions about how much  we understand about gravity in the weak and
    strong coupling regime.
And whether we can learn about gravitational physics beyond Einstein gravity,
like modified gravity scenarii (extra dimensions,
  massive gravity, \dots), or   quantum effects.

In this context it is important to work with  gravity effective field theories.
Although the status of  the  high-energy behaviour
of quantum gravity is still open, considering effective field theory of gravity at
low energy does not pose a problem. One can safely extract low-energy physics
from the quantization of the gravitational interactions observables
that are independent of the high-energy behaviour.  There are the
classical contributions (post-Minkowskian expansion) but as well
infra-red effects which are only sensitive to low-energy degrees of freedom.
As argued by J. D. Bjorken~\cite{Bjorken:2000zz}  
  \textsl{ [...] as an open theory,
    quantum gravity is arguably our best quantum field theory, not the
    worst.}

\section{Classical gravity from quantum $S$-matrix}
  
Einstein's theory of  gravity  is the first term of an effective field
theory coupling gravity to matter~\cite{Donoghue:1993eb,Donoghue:2017pgk}
\begin{equation}
{\cal S}_{EH} = \int d^4 x \sqrt{-g} \Bigg[\frac{R}{16 \pi G_N}  +
 g^{\mu\nu} T_{\mu\nu}^{\rm matter}+ \mathcal L_{\rm corrections}\Bigg]\,.
\end{equation} 
We assume that the effective field theory satisfies the standard
requierements of  locality, unitarity and Lorentz
invariance, and of course that the theory is diffeomorphism invariant
(i.e. we have the symmetries of General relativity).  The low-energy degrees of freedom are the massless
graviton and the  usual massive  matter fields.

We are interested in extracting physical observables from the
gravitational interactions\footnote{One could
as well include electro-magnetic interactions as considered in~\cite{Donoghue:2001qc} but
in this text we will only consider the exchange of  the graviton  and
focus on the gravitational sector.} between two massive body of masses $m_i$ and spin $S_i$ with $i=1,2$ interacting via the exchange of massless spin-2 graviton~\cite{tHooft:1974toh,Veltman:1975vx,DeWitt:1967yk,DeWitt:1967ub,DeWitt:1967uc}.
The two-body scattering matrix
\begin{equation}
{\mathcal M}(p_1,p_2,p_1',p_2')=\!\!\!\!\begin{gathered}
    \includegraphics[width=1.5cm]{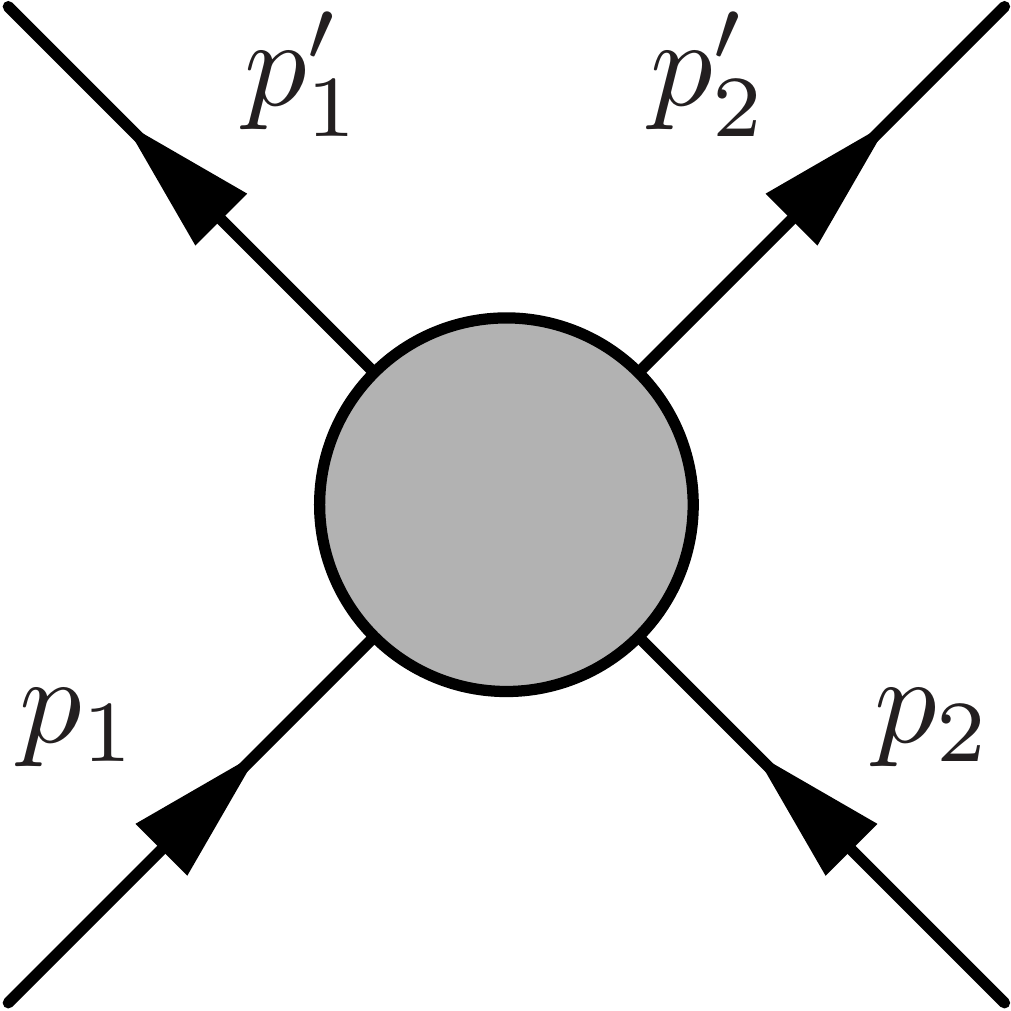}
  \end{gathered}
\!\!\!\!=\sum_{L\,=\,0}^{ \infty} \mathcal{M}_{ L}(p_1,p_2,p_1',p_2') ,
\end{equation}
can be expanded in perturbation $\mathcal{M}_{ L} \sim {\cal
  O}(G_N^{L+1})$.
The quantum scattering matrix ${\mathcal M}(p_1,p_2,p_1',p_2')$
depends on the incoming energy $p_1\cdot p_2/(m_1m_2)$, the
momentum transfer $(p_1-p_1')^2=q^2$ and $\hbar$.
At a given order in perturbation one gets the exchange of gravitons
(curly lines) between massive external matters (solid lines)
$$\includegraphics[width=11cm]{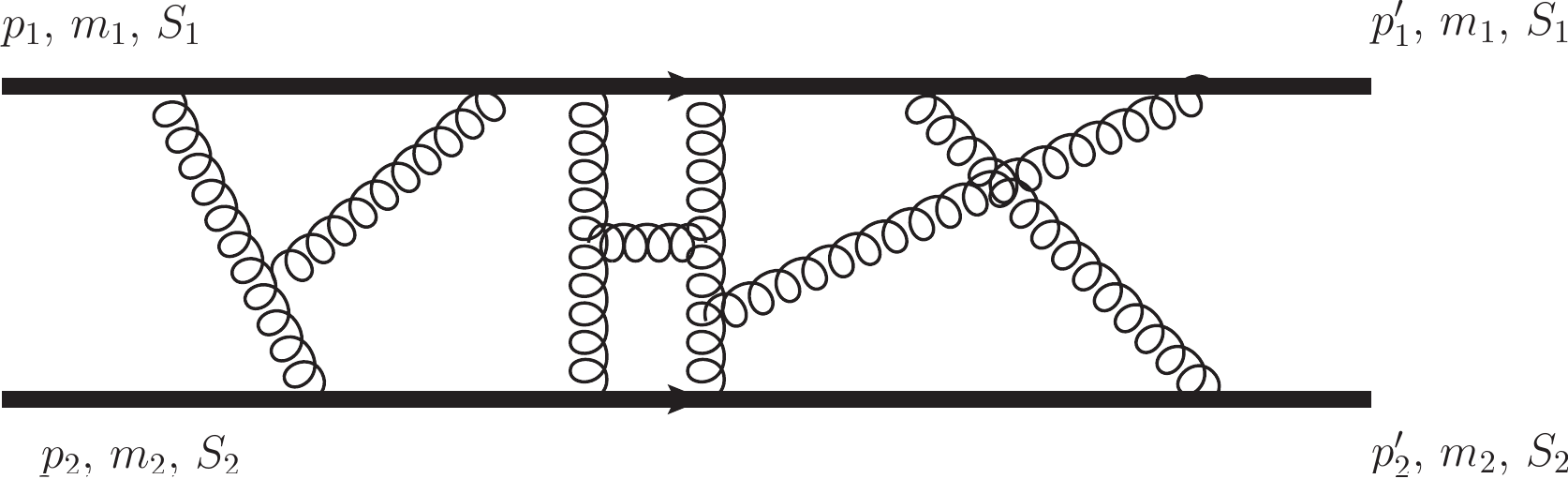}$$
From the limit $\hbar \to
    0$ with $q^2/\hbar$ held fixed of the quantum scattering amplitudes~\cite{Holstein:2004dn,Bjerrum-Bohr:2018xdl,Kosower:2018adc} , one can extract the classical  Hamiltonian $\mathcal{H}_{\rm
  PM}(p,r)$ for the gravitational  interactions between two  classical
massive body  in general
dimensions
\begin{equation}
\lim_{"\hbar\to0"}  \begin{gathered}
     \includegraphics[width=1.5cm]{2body.pdf}
   \end{gathered}
   =\sqrt{p^2+m_1^2}+\sqrt{p^2+m_2^2}
 +\sum_{L\geq0} G_N^{L+1} V_{\rm L+1-PM}(p,r).
 \end{equation}
The relativistic potential $G_N^{L+1}V_{\rm L+1-PM}(p,r)=\sum_{n\geq0}
c_{L,n}  G_N^{L+1} v^n$, with $p=p_1-p_1'=(E,v)$, resums an infinite
number of post-Newtonian contributions  in the small velocity
expansion $v/c\ll1$. The approach described here
works in any space-time dimension and applies to gravity in higher
dimensions~\cite{KoemansCollado:2019ggb,Cristofoli:2020uzm}. We will see an illustration of this when discussing the
black hole metric in section~\ref{sec:schw}.

  By computing the two-body scattering in perturbation one derives a
 Lorentz invariant expression valid in all regime of relative velocity
 between the two interacting massive bodies.   The
 scattering amplitude approach completes the post-Newtonian
 computations by providing information beyond its regime of validity,
 and leads to  surprising
 results connecting the conservative part  and
 gravitational radiation
 effects~\cite{Damour:2017zjx,Damour:2019lcq,Damour:2020tta,Parra-Martinez:2020dzs,DiVecchia:2021ndb,DiVecchia:2021bdo,Herrmann:2021tct,Bjerrum-Bohr:2021vuf}.  It gives a new perspective on the traditional
 methods~\cite{Goldberger:2007hy,Blanchet:2013haa,Porto:2016pyg,Barack:2018yly,Isoyama:2020lls}
 used for computing the gravitational wave templates.
This approach allows to connect the resummed
 post-Newtonian results~\cite{Bern:2019nnu,Bern:2019crd,Bern:2021dqo}
 and the high-energy
 behaviour~\cite{DiVecchia:2020ymx,DiVecchia:2021bdo}.

\section{The Schwarzschild-Tangherlini metric from scalar field
  amplitudes}
\label{sec:schw}
The Nobel prize citation for Roger Penrose states that ``black hole formation is a robust prediction of the general theory
    of relativity''. Subrahmanyan Chandrasekhar explained that they are the most perfect macroscopic objects there
    are in the universe since  the
only elements in their construction are our concepts of space and
time. Black-hole solutions are a perfect play ground to validate
the formalism of deriving classical gravity from quantum scattering amplitudes. This
 also opens new avenues for studying black holes in generalized
theories of gravity.

In 1973 Duff analyzed~\cite{Duff:1973zz}   the question of the classical limit of quantum
gravity by extracting the Schwarzschild back hole metric
from quantum tree graphs to $G_N^3$ order. This was a consistency
check on the way classical Einstein's gravity is embedded into the
standard massless spin-2 quantization of the gravitational interactions.

By evaluating the vertex function of the emission of a graviton from a
particle of mass $m$,  spin $S$ and charge $Q$, in $d$ dimensions
\begin{equation}\label{e:M3pt}
\begin{gathered}
  \includegraphics[]{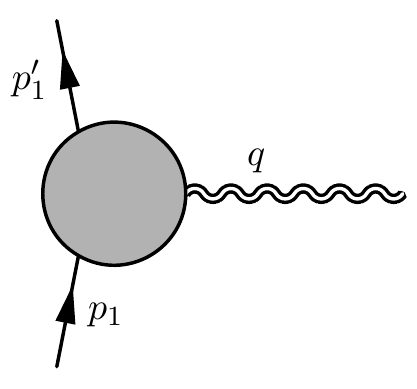}
\end{gathered}=-{i\sqrt{32\pi G_N}\over2}
\sum_{l\geq0}\langle  T^{(l)\, \mu\nu}(q^2)  \rangle\epsilon_{\mu\nu}
\end{equation}
with the action
\begin{equation}
\mathcal{S}=\int d^{d+1}x \sqrt{-g}\left({R\over 16\pi G_N}+
\frac{1}{2} g^{\mu\nu}\partial_{\mu}\phi \partial_{\nu}\phi-\frac{1}{2}m^2\phi^2\right)\,.
\end{equation}
one can extract the metric of   physical black holes~\cite{Vaidya:2014kza}
\begin{itemize}
\item Schwarzschild black hole: Scalar field $S=0$, mass $m$~\cite{Neill:2013wsa,Goldberger:2004jt,BjerrumBohr:2002ks,Mougiakakos:2020laz}
  \item Reissner-Nordstr\"om black hole: Scalar field  $S=0$, charge
    $Q$, mass $m$~\cite{Donoghue:2001qc}
 \item Kerr-Newman  black hole: Fermionic field  $S=\frac12$, charge
    $Q$, mass $m$~\cite{BjerrumBohr:2002ks,Donoghue:2001qc}
\end{itemize}
At each loop order we  extract the $l$-loop contribution to the
transition density   of the stress-energy tensor
$\langle T_{\mu\nu}(q^2)\rangle=\sum_{l\geq0} \langle T^{(l)}_{\mu\nu}(q^2)\rangle$
\begin{equation}\label{e:MtoT}
  i\mathcal M^{ (l )}_3(p_1,q)  =-{i\sqrt{32\pi G_N}\over2}
\langle  T^{(l)\, \mu\nu}(q^2)  \rangle\epsilon_{\mu\nu},
\end{equation}
where $\epsilon^{\mu\nu}$ is the polarisation of the graviton with
momentum $q=p_1-p_2$ is the   momentum transfer.

The  scattering amplitude computation is not done in the harmonic  gauge
coordinates $g^{\mu\nu}\Gamma^\lambda_{\mu\nu}(g)=0$  but in  the \textit{de
  Donder gauge}
coordinate system~\cite{Veltman:1975vx,Goldberger:2004jt,Cheung:2020gyp,Collado:2018isu,Jakobsen:2020ksu}

\begin{equation}\label{e:deDonderGauge}
{\eta^{\mu\nu}}\Gamma^\lambda_{\mu\nu}(g)= {\eta^{\mu\nu}}g^{\lambda\rho}\left({\partial g_{\rho\mu}\over \partial x^\nu}+ {\partial g_{\rho\nu}\over \partial x^\mu}-{\partial g_{\mu\nu}\over \partial x^\rho}\right)=0
\end{equation}
the metric  perturbations $g_{\mu\nu}=\eta_{\mu\nu} +\sum_{n\geq1}
h^{(n)}_{\mu\nu}$ satisfy\footnote{The   harmonic gauge linearized
  at the first order in perturbation gives~(\ref{e:hndD}) with
  $n=1$. The higher-order expansions of the harmonic
gauge differ from these conditions.}
\begin{equation}\label{e:hndD}
    {\partial\over
  \partial x^{\lambda}} h^{\lambda (n)}_{\nu}-\frac{1}{2} {\partial\over
  \partial x^{\nu}} h^{(n)}         =0\,.
\end{equation}
 The de Donder gauge relation between the
metric perturbation and the stress-energy tensor reads
\begin{equation}\label{e:TtohAmplitudedeDonder}
 h^{(l+1)}_{\mu\nu}(\vec x) = -16\pi G_N\int {d^d{\vec q}\over(2\pi)^d} e^{i\vec q\cdot \vec x} {1\over
 \vec q^2} \left( \langle T_{\mu\nu}^{(l)}\rangle^{\rm
   class.}(q^2)-\frac{1}{d-1}\eta_{\mu\nu}\langle T^{(l)}\rangle^{\rm class.}(q^2)\right)\,.
\end{equation}
In this relation enters the classical contribution at $l$ loop order $\langle T^{(l)}_{\mu\nu}\rangle^{\rm class.}(q^2) $ defined
by the classical limit of the quantum scattering
amplitude~\cite{Holstein:2004dn,Bjerrum-Bohr:2018xdl,Kosower:2018adc}
and its application to black hole solutions~\cite{Maybee:2019jus}.

In~\cite{Mougiakakos:2020laz}, the Schwarzschild metric up to $G_N^4$
obtained in four ($d=3$), five ($d=4$) and six ($d=5$) dimensions.
The  Schwarzschild-Tangherlini metric in the de donder coordinate
system
\begin{equation}
ds^2=\left(1-4{d-2\over d-1}\, {\rho(r,d)\over {   f(r)}^{d-2}}\right) dt^2-   {   f(r)}^2 d\vec x^2-\left(-{   f(r)}^2-{   f(r)}^{d-2} {({   f(r)}+r{d {   f(r)}\over dr})^2\over
                  {   f(r)}^{d-2}-4{d-2\over d-1}\rho(r,d) }\right) {(\vec           x\cdot
  d\vec x)^2\over \vec x^2}
\end{equation}
and to the first order the components of the metric are given~\cite{Mougiakakos:2020laz} In four
dimensions ($d=3$)\footnote{Similar results are obtained in higher
dimensions where we matched the
Schwarzschild-Tangherlini up to the order $G_N^4$.}
{\small\begin{equation}\label{e:h0finite}
  h^{\rm dD}_0(r)=1-\frac{ 2G_N m}{r}+2\left(\frac{ G_N
      m}{r}\right)^2+2\left(\frac{ G_N m}{r}\right)^3
  +\left(\frac43 { \log
      \left(\frac{r C_3}{G_N m}\right)}-6\right) \left(G_N m\over
    r\right)^4+\cdots
  \end{equation}}
{\small\begin{multline*}\label{e:h1finite}
  h^{\rm dD}_1(r)=1+2\frac{ G_N m}{r}+5\left(\frac{ G_N m}{r}\right)^2+
    \left({4\over3}{  \log
      \left(\frac{rC_3}{G_N m}\right)}+ 4\right)\left(G_N m\over
      r\right)^3\cr
    +
   \left(-\frac43 \log
      \left(\frac{rC_3}{G_N m}\right) +\frac{16}{3}\right)\left(G_N m\over r\right)^4
      +  \left(\frac{64}{15} \log
      \left(\frac{rC_3}{G_N m}\right) -\frac{26}{75}\right)\left(G_N m\over r\right)^5\cr
       +\Bigg(\frac{4}{9} { \log
      \left(\frac{rC_3}{G_N m}\right)^2}-\frac{24}{5}\log
      \left(\frac{rC_3}{G_N m}\right) 
     +\frac{298}{75}\Bigg) \left(G_N m\over
      r\right)^6+\cdots
\end{multline*}}
One notices that in this gauge the metric components contain
finite-size    powers of $\log(r C_3/G_Nm)$ where $C_3$ is the single
constant of integration.
These logarithms are generated by the cancellation of the  ultraviolet divergences
 of the scattering amplitudes in~\eqref{e:MtoT} regulated by the
 introduction of higher-derivative non-minimal couplings~\cite{Goldberger:2004jt,Jakobsen:2020ksu,Mougiakakos:2020laz}. These
contributions are finite size effects~\cite{Goldberger:2004jt,Foffa:2013qca,Porto:2016pyg,Huber:2020xny,Cheung:2020sdj,Haddad:2020que,Bern:2020uwk,Kalin:2020lmz}. At the level of the
metric components, they are reabsorbed in
the coordinate change from the de Donder gauge used for the amplitude
computation to the standard Schwarzschild-Tangherlini metric in
spherical coordinates
\begin{equation}
  \label{eq:ScTspherical}
  ds^2=\left(1- 4 {d-2\over d-1}\rho(r,d) \right) dt^2 - d{\vec x}^2-
  {4{d-2\over d-1}\rho(r,d)\over 1-4 {d-2\over d-1}\rho(r,d)} {(\vec
    x\cdot d\vec x)^2\over r^2}; \qquad
  \rho(r,d)={\Gamma\left(d-2\over2\right) \over \pi^{d-2\over2}}
  {G_Nm\over r^{d-2}}\,,
\end{equation}
and the finite-size effects do not affect the static metric.

\section{Discussion} \label{sec:discussion}

General relativity can be considered in  space-times of various
dimensions. Gravity is richer in higher dimensions as black-hole solutions develop non trivial properties in general
dimensions~\cite{Emparan:2008eg,Emparan:2013moa}. It is therefore important to validate our current
understanding of the connection between the quantum scattering amplitudes and
classical general relativity in general
dimensions~\cite{KoemansCollado:2019ggb,Cristofoli:2020uzm}.
By reproducing the classical
Schwarzschild-Tangherlini  metric from scattering amplitudes in four,
five  and six dimensions,  we validate the procedure for
extracting the classical piece from the quantum scattering amplitudes.
 The method can be applied to derive other  black-hole metrics, like the Kerr-Newman and Reissner-Nordstr\"om metrics~ by 
considering  the vertex function of the
emission of the graviton from a massive particle with  spin and charge.~\cite{Donoghue:2001qc,BjerrumBohr:2002ks,Guevara:2018wpp,Chung:2018kqs,Moynihan:2019bor,Chung:2019yfs,Guevara:2019fsj,Cristofoli:2020hnk}.

The scattering amplitude approach presented in this work can be applied to
any effective field theory of gravity coupled to matter fields.
One can include quantum corrections in~\eqref{e:TtohAmplitudedeDonder}
and examine the impact of quantum effects on the black hole
solution~\cite{BjerrumBohr:2002ks}.  Or study the impact of higher
derivative contributions to the gravitational waves templates.

The amplitudes computations, being performed in general
dimensions, lead to results that have an analytic dependence on the
space-time dimensions.

The higher order post-Minkowskian contributions should be obtained from
higher-loop amplitudes in a direct application of the methods
described here. Up to the third Post-Minkowskian order one can extract
the classical conservative potential and scattering angle of binary
system in pure gravity and maximal supergravity~\cite{Bjerrum-Bohr:2013bxa,Bern:2019nnu,Bern:2019crd,Bern:2021dqo,Parra-Martinez:2020dzs,DiVecchia:2020ymx,DiVecchia:2021bdo,Bjerrum-Bohr:2021vuf}.

 We have improved our understanding of the
  relation between general relativity and the quantum theory of
  gravity.  This leads to many new exciting developpements leading to a
  better understanding of the gravitational interactions in binary
  system.  This  provides  new techniques that can be applied to any gravitational
   effective field theories which have
    amplitude description : opening the possibility to search for
    deviation from Einstein gravity.

\section*{Acknowledgements}
We would like to thank Emil Bjerrum-Bohr, Poul Damgaard, Stavros
Mougiakakos and Ludovic Plant\'e for collaboration on the work
reported here.
The research of P. Vanhove has received funding from the ANR
grant ``Amplitudes'' ANR-17- CE31-0001-01, and the ANR grant ``SMAGP''
ANR-20-CE40-0026-01 and is partially supported by Laboratory of Mirror
Symmetry NRU HSE, RF Government grant, ag. No 14.641.31.0001.


\section*{References}

\begin{thebibliography}{99}


\bibitem{Abbott:2016blz}
  B.~P.~Abbott \textit{et al.} [LIGO Scientific and Virgo],
 ``Observation of Gravitational Waves from a Binary Black Hole Merger,''
Phys. Rev. Lett. \textbf{116} (2016) no.6, 061102
[arXiv:1602.03837 [gr-qc]].

\bibitem{TheLIGOScientific:2017qsa}
B.~P.~Abbott \textit{et al.} [LIGO Scientific and Virgo],
``GW170817: Observation of Gravitational Waves from a Binary Neutron Star Inspiral,''
Phys. Rev. Lett. \textbf{119} (2017) no.16, 161101
[arXiv:1710.05832 [gr-qc]].

  
\bibitem{Barausse:2020rsu}
E.~Barausse, E.~Berti, T.~Hertog, S.~A.~Hughes, P.~Jetzer, P.~Pani, T.~P.~Sotiriou, N.~Tamanini, H.~Witek and K.~Yagi, \textit{et al.}
``Prospects for Fundamental Physics with Lisa,''
Gen. Rel. Grav. \textbf{52} (2020) no.8, 81
doi:10.1007/s10714-020-02691-1
[arXiv:2001.09793 [gr-qc]].


\bibitem{Bjorken:2000zz}
J.~D.~Bjorken,
``The Future of Particle Physics,''
Int. J. Mod. Phys. A \textbf{16} (2001), 483-502
[arXiv:hep-ph/0006180 [hep-ph]].

\bibitem{Donoghue:1993eb}
J.~F.~Donoghue,
``Leading Quantum Correction to the Newtonian Potential,''
Phys. Rev. Lett. \textbf{72} (1994), 2996-2999
[arXiv:gr-qc/9310024 [gr-qc]].

\bibitem{Donoghue:2017pgk}
J.~F.~Donoghue, M.~M.~Ivanov and A.~Shkerin,
``Epfl Lectures on General Relativity as a Quantum Field Theory,''
[arXiv:1702.00319 [hep-th]].

\bibitem{Donoghue:2001qc}
J.~F.~Donoghue, B.~R.~Holstein, B.~Garbrecht and T.~Konstandin,
``Quantum Corrections to the Reissner-Nordstr\"Om and Kerr-Newman Metrics,''
Phys. Lett. B \textbf{529} (2002), 132-142
[erratum: Phys. Lett. B \textbf{612} (2005), 311-312]
[arXiv:hep-th/0112237 [hep-th]].

  
\bibitem{tHooft:1974toh}
G.~'t Hooft and M.~J.~G.~Veltman,
``One Loop Divergencies in the Theory of Gravitation,''
Ann. Inst. H. Poincare Phys. Theor. A \textbf{20} (1974), 69-94


\bibitem{Veltman:1975vx}
  M.~J.~G.~Veltman,
  ``Quantum Theory of Gravitation,''
  Conf.\ Proc.\ C {\bf 7507281} (1975) 265.


\bibitem{DeWitt:1967yk}
B.~S.~DeWitt,
``Quantum Theory of Gravity. 1. The Canonical Theory,''
Phys. Rev. \textbf{160} (1967), 1113-1148


\bibitem{DeWitt:1967ub}
B.~S.~DeWitt,
``Quantum Theory of Gravity. 2. The Manifestly Covariant Theory,''
Phys. Rev. \textbf{162} (1967), 1195-1239

\bibitem{DeWitt:1967uc}
B.~S.~DeWitt,
``Quantum Theory of Gravity. 3. Applications of the Covariant Theory,''
Phys. Rev. \textbf{162} (1967), 1239-1256

\bibitem{Neill:2013wsa}
D.~Neill and I.~Z.~Rothstein,
``Classical Space-Times from the S Matrix,''
Nucl. Phys. B \textbf{877} (2013), 177-189
[arXiv:1304.7263 [hep-th]].


\bibitem{Holstein:2004dn}
B.~R.~Holstein and J.~F.~Donoghue,
``Classical Physics and Quantum Loops,''
Phys. Rev. Lett. \textbf{93} (2004), 201602
[arXiv:hep-th/0405239 [hep-th]].

\bibitem{Bjerrum-Bohr:2018xdl}
N.~E.~J.~Bjerrum-Bohr, P.~H.~Damgaard, G.~Festuccia, L.~Plant\'e and P.~Vanhove,
``General Relativity from Scattering Amplitudes,''
Phys. Rev. Lett. \textbf{121} (2018) no.17, 171601
[arXiv:1806.04920 [hep-th]].

  
\bibitem{Kosower:2018adc}
D.~A.~Kosower, B.~Maybee and D.~O'Connell,
``Amplitudes, Observables, and Classical Scattering,''
JHEP \textbf{02} (2019), 137
[arXiv:1811.10950 [hep-th]].

  
\bibitem{KoemansCollado:2019ggb}
A.~Koemans Collado, P.~Di Vecchia and R.~Russo,
``Revisiting the second post-Minkowskian eikonal and the dynamics of binary black holes,''
Phys. Rev. D \textbf{100} (2019) no.6, 066028
[arXiv:1904.02667 [hep-th]].

\bibitem{Cristofoli:2020uzm}
A.~Cristofoli, P.~H.~Damgaard, P.~Di Vecchia and C.~Heissenberg,
``Second-order Post-Minkowskian scattering in arbitrary dimensions,''
JHEP \textbf{07} (2020), 122
[arXiv:2003.10274 [hep-th]].

\bibitem{Damour:2017zjx}
T.~Damour,
``High-Energy Gravitational Scattering and the General Relativistic Two-Body Problem,''
Phys. Rev. D \textbf{97} (2018) no.4, 044038
[arXiv:1710.10599 [gr-qc]].

\bibitem{Damour:2019lcq}
T.~Damour,
``Classical and Quantum Scattering in Post-Minkowskian Gravity,''
Phys. Rev. D \textbf{102} (2020) no.2, 024060
[arXiv:1912.02139 [gr-qc]].

\bibitem{Damour:2020tta}
T.~Damour,
``Radiative Contribution to Classical Gravitational Scattering at the Third Order in $G$,''
Phys. Rev. D \textbf{102} (2020) no.12, 124008
[arXiv:2010.01641 [gr-qc]].

\bibitem{Parra-Martinez:2020dzs}
J.~Parra-Mart{\'\i ne}z, M.~S.~Ruf and M.~Zeng,
``Extremal Black Hole Scattering at $\mathcal{O}(G^3)$: Graviton Dominance, Eikonal Exponentiation, and Differential Equations,''
JHEP \textbf{11} (2020), 023
[arXiv:2005.04236 [hep-th]].

\bibitem{DiVecchia:2021ndb}
P.~Di Vecchia, C.~Heissenberg, R.~Russo and G.~Veneziano,
``Radiation Reaction from Soft Theorems,''
[arXiv:2101.05772 [hep-th]].

\bibitem{DiVecchia:2021bdo}
P.~Di Vecchia, C.~Heissenberg, R.~Russo and G.~Veneziano,
``The Eikonal Approach to Gravitational Scattering and Radiation at $\mathcal O(G^3)$,''
[arXiv:2104.03256 [hep-th]].


\bibitem{Herrmann:2021tct}
E.~Herrmann, J.~Parra-Mart{\'\i ne}z, M.~S.~Ruf and M.~Zeng,
``Radiative Classical Gravitational Observables at $\mathcal O(G^3)$ from Scattering Amplitudes,''
[arXiv:2104.03957 [hep-th]].

  
\bibitem{Bjerrum-Bohr:2021vuf}
N.~E.~J.~Bjerrum-Bohr, P.~H.~Damgaard, L.~Plant\'e and P.~Vanhove,
``Classical Gravity from Loop Amplitudes,''
[arXiv:2104.04510 [hep-th]].

\bibitem{Goldberger:2007hy}
W.~D.~Goldberger,
``Les Houches Lectures on Effective Field Theories and Gravitational Radiation,''
[arXiv:hep-ph/0701129 [hep-ph]].

\bibitem{Blanchet:2013haa}
L.~Blanchet,
``Gravitational Radiation from Post-Newtonian Sources and Inspiralling Compact Binaries,''
Living Rev. Rel. \textbf{17} (2014), 2
[arXiv:1310.1528 [gr-qc]].

\bibitem{Porto:2016pyg}
R.~A.~Porto,
``The Effective Field Theorist's Approach to Gravitational Dynamics,''
Phys. Rept. \textbf{633} (2016), 1-104
[arXiv:1601.04914 [hep-th]].

\bibitem{Barack:2018yly}
L.~Barack, V.~Cardoso, S.~Nissanke, T.~P.~Sotiriou, A.~Askar, C.~Belczynski, G.~Bertone, E.~Bon, D.~Blas and R.~Brito, \textit{et al.}
``Black Holes, Gravitational Waves and Fundamental Physics: a Roadmap,''
Class. Quant. Grav. \textbf{36} (2019) no.14, 143001
[arXiv:1806.05195 [gr-qc]].
  
\bibitem{Isoyama:2020lls}
S.~Isoyama, R.~Sturani and H.~Nakano,
``Post-Newtonian Templates for Gravitational Waves from Compact Binary Inspirals,''
[arXiv:2012.01350 [gr-qc]].

\bibitem{Bern:2019nnu}
Z.~Bern, C.~Cheung, R.~Roiban, C.~H.~Shen, M.~P.~Solon and M.~Zeng,
``Scattering Amplitudes and the Conservative Hamiltonian for Binary Systems at Third Post-Minkowskian Order,''
Phys. Rev. Lett. \textbf{122} (2019) no.20, 201603
[arXiv:1901.04424 [hep-th]].

\bibitem{Bern:2019crd}
Z.~Bern, C.~Cheung, R.~Roiban, C.~H.~Shen, M.~P.~Solon and M.~Zeng,
``Black Hole Binary Dynamics from the Double Copy and Effective Theory,''
JHEP \textbf{10} (2019), 206
[arXiv:1908.01493 [hep-th]].

\bibitem{Bern:2021dqo}
Z.~Bern, J.~Parra-Mart{\'\i ne}z, R.~Roiban, M.~S.~Ruf, C.~H.~Shen, M.~P.~Solon and M.~Zeng,
``Scattering Amplitudes and Conservative Binary Dynamics at ${\cal O}(G^4)$,''
[arXiv:2101.07254 [hep-th]].

\bibitem{DiVecchia:2020ymx}
P.~Di Vecchia, C.~Heissenberg, R.~Russo and G.~Veneziano,
``Universality of Ultra-Relativistic Gravitational Scattering,''
Phys. Lett. B \textbf{811} (2020), 135924
[arXiv:2008.12743 [hep-th]].

\bibitem{Duff:1973zz}
M.~J.~Duff,
``Quantum Tree Graphs and the Schwarzschild Solution,''
Phys. Rev. D \textbf{7} (1973), 2317-2326

\bibitem{BjerrumBohr:2002ks}
N.~E.~J.~Bjerrum-Bohr, J.~F.~Donoghue and B.~R.~Holstein,
``Quantum Corrections to the Schwarzschild and Kerr Metrics,''
Phys. Rev. D \textbf{68} (2003), 084005
[erratum: Phys. Rev. D \textbf{71} (2005), 069904]
[arXiv:hep-th/0211071 [hep-th]].


\bibitem{Mougiakakos:2020laz}
S.~Mougiakakos and P.~Vanhove,
``Schwarzschild-Tangherlini Metric from Scattering Amplitudes in Various Dimensions,''
Phys. Rev. D \textbf{103} (2021) no.2, 026001
[arXiv:2010.08882 [hep-th]].

\bibitem{Goldberger:2004jt}
W.~D.~Goldberger and I.~Z.~Rothstein,
``An Effective Field Theory of Gravity for Extended Objects,''
Phys. Rev. D \textbf{73} (2006), 104029
[arXiv:hep-th/0409156 [hep-th]].



\bibitem{Cheung:2020gyp}
C.~Cheung and M.~P.~Solon,
``Classical gravitational scattering at $ \mathcal{O} (G^{3})$ from Feynman diagrams,''
JHEP \textbf{06} (2020), 144
[arXiv:2003.08351 [hep-th]].

\bibitem{Collado:2018isu}
A.~K.~Collado, P.~Di Vecchia, R.~Russo and S.~Thomas,
``The subleading eikonal in supergravity theories,''
JHEP \textbf{10} (2018), 038
[arXiv:1807.04588 [hep-th]].

  
\bibitem{Jakobsen:2020ksu}
G.~U.~Jakobsen,
``Schwarzschild-Tangherlini Metric from Scattering Amplitudes,''
Phys. Rev. D \textbf{102} (2020) no.10, 104065
[arXiv:2006.01734 [hep-th]].

\bibitem{Foffa:2013qca}
S.~Foffa and R.~Sturani,
``Effective Field Theory Methods to Model Compact Binaries,''
Class. Quant. Grav. \textbf{31} (2014) no.4, 043001
[arXiv:1309.3474 [gr-qc]].



\bibitem{Huber:2020xny}
M.~Accettulli Huber, A.~Brandhuber, S.~De Angelis and G.~Travaglini,
``From Amplitudes to Gravitational Radiation with Cubic Interactions and Tidal Effects,''
Phys. Rev. D \textbf{103} (2021) no.4, 045015
[arXiv:2012.06548 [hep-th]].

\bibitem{Cheung:2020sdj}
C.~Cheung and M.~P.~Solon,
``Tidal Effects in the Post-Minkowskian Expansion,''
Phys. Rev. Lett. \textbf{125} (2020) no.19, 191601
[arXiv:2006.06665 [hep-th]].

\bibitem{Haddad:2020que}
K.~Haddad and A.~Helset,
``Tidal Effects in Quantum Field Theory,''
JHEP \textbf{12} (2020), 024
[arXiv:2008.04920 [hep-th]].

\bibitem{Bern:2020uwk}
Z.~Bern, J.~Parra-Mart{\'\i ne}z, R.~Roiban, E.~Sawyer and C.~H.~Shen,
``Leading Nonlinear Tidal Effects and Scattering Amplitudes,''
[arXiv:2010.08559 [hep-th]].

\bibitem{Kalin:2020lmz}
G.~K\"alin, Z.~Liu and R.~A.~Porto,
``Conservative Tidal Effects in Compact Binary Systems to Next-To-Leading Post-Minkowskian Order,''
Phys. Rev. D \textbf{102} (2020), 124025
[arXiv:2008.06047 [hep-th]].

\bibitem{Emparan:2008eg}
R.~Emparan and H.~S.~Reall,
``Black Holes in Higher Dimensions,''
Living Rev. Rel. \textbf{11} (2008), 6
[arXiv:0801.3471 [hep-th]].

\bibitem{Emparan:2013moa}
R.~Emparan, R.~Suzuki and K.~Tanabe,
``The large D limit of General Relativity,''
JHEP \textbf{06} (2013), 009
[arXiv:1302.6382 [hep-th]].

\bibitem{Guevara:2018wpp}
A.~Guevara, A.~Ochirov and J.~Vines,
``Scattering of Spinning Black Holes from Exponentiated Soft Factors,''
JHEP \textbf{09} (2019), 056
[arXiv:1812.06895 [hep-th]].

\bibitem{Chung:2018kqs}
M.~Z.~Chung, Y.~T.~Huang, J.~W.~Kim and S.~Lee,
``The simplest massive S-matrix: from minimal coupling to Black Holes,''
JHEP \textbf{04} (2019), 156
[arXiv:1812.08752 [hep-th]].

\bibitem{Moynihan:2019bor}
N.~Moynihan,
``Kerr-Newman from Minimal Coupling,''
JHEP \textbf{01} (2020), 014
[arXiv:1909.05217 [hep-th]].

\bibitem{Chung:2019yfs}
M.~Z.~Chung, Y.~T.~Huang and J.~W.~Kim,
``Kerr-Newman stress-tensor from minimal coupling,''
JHEP \textbf{12} (2020), 103
[arXiv:1911.12775 [hep-th]].

\bibitem{Guevara:2019fsj}
A.~Guevara, A.~Ochirov and J.~Vines,
``Black-hole scattering with general spin directions from minimal-coupling amplitudes,''
Phys. Rev. D \textbf{100} (2019) no.10, 104024
[arXiv:1906.10071 [hep-th]].

\bibitem{Cristofoli:2020hnk}
A.~Cristofoli,
``Gravitational shock waves and scattering amplitudes,''
JHEP \textbf{11} (2020), 160
[arXiv:2006.08283 [hep-th]].

\bibitem{Bjerrum-Bohr:2013bxa}
N.~E.~J.~Bjerrum-Bohr, J.~F.~Donoghue and P.~Vanhove,
``On-Shell Techniques and Universal Results in Quantum Gravity,''
JHEP \textbf{02} (2014), 111
[arXiv:1309.0804 [hep-th]].

\bibitem{Vaidya:2014kza}
V.~Vaidya,
``Gravitational spin Hamiltonians from the S matrix,''
Phys. Rev. D \textbf{91} (2015) no.2, 024017;
[arXiv:1410.5348 [hep-th]].

\bibitem{Maybee:2019jus}
B.~Maybee, D.~O'Connell and J.~Vines,
``Observables and amplitudes for spinning particles and black holes,''
JHEP \textbf{12} (2019), 156;
[arXiv:1906.09260 [hep-th]].


\end{thebibliography}
\end{document}